\documentclass[a4paper]{jpconf}
\usepackage{graphicx}
\begin{document}
\title{Technical aspects and dark matter searches}

\author{R Bernabei $^{1,2}$, P Belli $^{2}$, F Cappella $^{3,4}$,
R Cerulli $^{5}$, C J Dai $^{6}$,
A d'Angelo $^{3,4}$, H L He $^{6}$, A Incicchitti $^{4}$, H H Kuang $^{6}$,
X H Ma $^{6}$, F Montecchia $^{2,7}$, F Nozzoli $^{1,2}$,
D Prosperi$^{3,4}$, X D Sheng$^{6}$ and Z P Ye$^{6,8}$}

\address{$^1$ Dip. di Fisica, Universit\`a di Roma ``Tor Vergata'', I-00133 Rome, Italy}

\address{$^2$ INFN, sez. Roma ``Tor Vergata'', I-00133 Rome, Italy}

\address{$^3$ Dip. di Fisica, Universit\`a di Roma ``La Sapienza'', I-00185 Rome, Italy}

\address{$^4$ INFN, sez. Roma, I-00185 Rome, Italy}

\address{$^5$ Laboratori Nazionali del Gran Sasso, I.N.F.N., Assergi, Italy}

\address{$^6$ IHEP, Chinese Academy, P.O. Box 918/3, Beijing 100039, China}

\address{$^7$ Laboratorio Sperimentale Policentrico di Ingegneria Medica, Universit\`a
degli Studi di Roma ``Tor Vergata''}

\address{$^8$ University of Jing Gangshan, Jiangxi, China}

\ead{rita.bernabei@roma2.infn.it}

\begin{abstract}
A variety of detectors has been proposed 
for dark matter direct detection, but most of them -- by the fact -- are still at R\&D stage.
In many cases, it is claimed that the lack of an adequate detectors' radio-purity 
might be compensated through heavy uses of MonteCarlo 
simulations, subtractions and handlings
of the measured counting rates, in order to claim higher sensitivity (just for a particular scenario).
The relevance of a correct evaluation of systematic effects in the use of MonteCarlo 
simulations at very low energy (which has always been safely discouraged in the field so far)
and of multiple subtractions and handling procedures applied to the measured counting rate is shortly
addressed here at some extent. Many other aspects would also deserve suitably deep investigations. 
\end{abstract}

In this paper some arguments presented at the TAUP09 conference will be shortly summarized. 
More details, tables and figures can be found in the slides at the conference site\cite{nozzoli}.

\vspace{0.3cm}

Let us firstly comment the possibility of reliable evaluations of the background contributions at the keV
energy region in the field of Dark Matter searches. As well known, it has been generally discouraged this 
procedure in the field of Dark Matter over more than twenty years.
In fact, the estimation by a MonteCarlo simulation of the background component in the counting rate from the residual radioactivity 
requires a detailed knowledge of: i) the exact set-up geometry 
(detector or detectors' matrix, all materials, details of the assembling, of the shield layers, of the site, etc.);
ii) the detector response function (e.g. energy resolution, $\alpha/\beta$ ratio, channeling, etc.); 
iii) the nature, the position and the concentration of all the existing radioactive contaminants;
iv) etc..
Unfortunately, apart from the geometrical layout of the set-up that are generally well known
by people inside the experimental group, all the other quantities necessary in the MonteCarlo
simulation require dedicated measurements. Moreover, there are some quantities
(such as concentration of residual contaminants, etc.) that can be poorly known and
just upper/lower limits are available; in some cases these quantities can be even totally unknown.
As an example, the experimental energy resolution as a function of the energy and the energy scale
should be measured/verified down to the energy threshold, as done e.g. by the DAMA/LIBRA experiment 
where they are continuously measured by external/internal known sources from MeV down to
the energy threshold \cite{perflibra}. On the contrary, in other experiments these quantities are instead extrapolated 
from calibrations at much higher energy (as done e.g. by liquid noble gas
set-ups, where the energy threshold and the few keV energy scale are 
generally unproven, also because of position dependence, of non-uniform signal collection, etc.; see e.g. \cite{paperliq}). 
Regarding the presence of residual contaminants in the set-up, 
generally only limits on the contributions of the ``standard'' contaminants 
are given; these limits forbid any reliable estimation of the background (being unknown the exact values)
and cannot be obviously exceeded (see also later). 
Moreover, possible presence of many non-standard contaminants should be also included.
In addition, the MonteCarlo simulation also depends on the precise 
location of all the contaminants -- that is generally unknown even for the ``standard'' ones -- in complex set-ups.
The situation is more complex for multi-detectors set-up and when the energy distribution refers to events where each 
detector has all the others in anticoincidence ({\it single-hit} events).
Thus, it is trivial to conclude that a reliable precise simulation of the background counting rate -- in particular at keV energy region -- 
is not univocally determined and is a quite impossible task.

In addition, beyond the fore-mentioned arguments, we need to take into account
that a MonteCarlo code cannot manage 
all the possible low energy atomic physical processes.
This argument is still subject of improvements; in fact,
as an example, non-negligible differences are also obtained
by different versions of the same MonteCarlo code \cite{geant4}.

Some instructive examples are given by the trails in MonteCarlo simulations in ref. \cite{kundryavtsev}.
As shown there, these simulations noticeably differ from the measured energy distributions in the cases of XENON-10 and of ZEPLIN-III.
In fact, they predict twice the measured rate for XENON-10 near 200 keV, and more than one order of magnitude 
the measured rate for ZEPLIN-III in the MeV range \cite{kundryavtsev}.
The same approach has been also pursued by the same authors \cite{kundryavtsev} trying
a MonteCarlo simulation of the background in the DAMA/LIBRA set-up. In particular, apart from 
errors in the details of the set-up geometry 
reconstruction and in the multiple-hit definition, many crude and arbitrary approximations 
in the nature and in the location of the residual contaminants
have been arbitrarily assumed; in fact, e.g.: i) only standard contaminants, ii) only unbroken chains, iii) only uniform 
location of contaminants in the detectors, etc. have been taken into account.
As a result of this rough, partial and arbitrary approach, the predicted rate has been 
estimated within a factor 10 lower than the measured one. Instead of refining
the quality of the simulation or reasonably recognizing the impossibility of precise determination, 
the authors just pursued the exercise of arbitrarily increasing ``by hand'' 
the assumed values of the contaminants at levels much larger 
than the measured experimental limits \cite{kundryavtsev}. This also implies
an overestimate of the background in higher energy region with respect to the measured 
experimental rate. In conclusion, although the arbitrary and the erroneous adopted procedures,
these authors do not succeed in reproducing either the
low or the high energy spectra. Nevertheless, 
as a conclusion of this arbitrary exercise \cite{kundryavtsev}, this artificially-boosted
simulated spectrum has been subtracted by the measured one, attempting to obtain a limit for
the unmodulated Dark Matter signal component.
This example shows how subtraction procedures using
MonteCarlo simulations in the few keV energy region 
can give rise to erroneous conclusions; thus, 
any constraint on Dark Matter signal on this basis would be an artefact.

Furthermore, let us also note that the measured spectra e.g. of the existing/past NaI(Tl) 
detectors (such as e.g. ANAIS, Frejus, NAIAD, ELEGANT, etc...) do not support even the shape
presented in ref. \cite{kundryavtsev}. 
In addition, well different counting rates at keV energy region
are present even for detectors of the same experimental group, 
as e.g. the case of NAIAD in 1996 and in 2003 \cite{nozzoli}.
In conclusion, it does not exist an unique recipe for a precise and reliable MonteCarlo simulation
of whatever set-up, and for NaI(Tl) in particular. 

Let us finally remind that a safer approach has been presented in this conference by DAMA collaboration;
this shows that enough space is present in the measured counting rate of the DAMA/LIBRA keV energy spectrum
for the unmodulated component of Dark Matter signal \cite{nozzoli,modlibra}.

\vspace{0.3cm}

In the second part of this contribution \cite{nozzoli},
problems related to the application of multiple subtraction procedures 
of the measured counting rate, as pursued by experiments trying to identify 
the presence of recoil nuclei in the measured energy spectrum, have been summarized.
In fact, many of the existing Dark Matter
candidates -- also in the WIMP class -- can give rise to signals 
that either have totally an electromagnetic nature (see e.g. \cite{ijma,wimpele,ldm}) 
or involve electromagnetic signals associated to 
nuclear recoils (see e.g. \cite{ijma07,chan}); 
obviously, approaches that are based on
multiple subtraction procedure of the electromagnetic component of the
counting rate are blind to similar scenarios. Moreover, well known side processes
exist for recoils (such as recoils induced by neutrons, fission fragments, end-range alphas, 
surface electrons, etc.).
This approach is generally pursued when the
detectors suffer from a not-suitable 
radiopurity level in the sensitive target-material and in the surroundings.

Those activities generally apply a large number of cut procedures
to the data; each one is affected by non-negligible systematic errors
which are usually not suitable quantified.
As an example, the XENON-10 experiment applies more than 10 different cuts
to the data; the experiment collects $\sim 10^4$ events
but only 10 are claimed to survive to the cuts and handling procedures \cite{xenon10}.
Very high reduction factors following applied multiple cuts are dangerous 
because of the difficult precise estimate of all the involved systematics.
For example, it has been shown in ref. \cite{benoit} that
ZEPLIN-I has claimed a sensitivity 3 orders of magnitude larger than the one 
properly obtained when accounting for systematics.
Thus, the robustness of some results appeared in the ``race for the best exclusion-plot'' 
(valid just in a single set of assumptions for a certain kind of WIMP) 
should be considered ``cum grano salis''.
For example, for some liquid noble gas set-ups,
apart from the robustness of the applied cuts themselves,
the very low energy scale and the energy threshold are 
determined by extrapolating from calibrations at much higher energy and 
applying some kind of corrections for relevant non-uniformity of the detector's response;
with a light yield of about 2.2 photoelectrons/keV and not specific calibrations even 
$\approx$ 1.5-2 keV electron equivalent is claimed as energy threshold \cite{xenon10,nozzoli}.
Considering the energy threshold dependence of the 
exclusion plots, up to several orders of
magnitude differences can be present between claimed and realistic
evaluation of the experimental sensitivity. 

Another crucial aspect is the proper accounting for the existing experimental and theoretical
uncertainties in the calculation and in the comparison of experiments using different target
materials and approaches.

\end{document}